\documentclass[conf]{vgtc}                          % final (conference style)
\graphicspath{{figures/}{pictures/}{images/}{./}} % where to search for the images

\usepackage{times}                     % we use Times as the main font
\usepackage{mathptmx}                  % use matching math font

\usepackage[inline]{enumitem}
\usepackage{amsmath}
\usepackage{amssymb}
\usepackage{mathtools}
\usepackage{hyphenat}
\usepackage{fnpct}
\usepackage{csquotes}
\usepackage{cite}  % sort citations
\usepackage[svgnames]{xcolor}  % import here for todonotes, so that the same arguments as in \vgtcinsertpkg are used
\usepackage[textsize=tiny]{todonotes}
\usepackage{tikz}
\usepackage{subcaption}

\usepackage[
  detect-all,
  round-pad=false,
  group-digits=integer,
  round-mode=places,
  round-precision=3,
]{siunitx}
\DeclareSIUnit\pixel{px}

\RequirePackage{expl3}
\RequirePackage{xparse}

\ExplSyntaxOn
\prop_new:N \l__citet_names_prop
\prop_clear:N \l__citet_names_prop

\cs_new:Nn \citet_addname:nn {
  \prop_put:Nnn \l__citet_names_prop {#1} {#2}
}

\cs_new:Nn \citet_usename:n {
  \prop_get:NnNTF \l__citet_names_prop {#1} \l_tmpa_tl {
    \l_tmpa_tl
  } {
    \PackageWarning{citet}{Citation~key~#1~was~not~found.}
    \textbf{??}
  }
}

\cs_new:Nn \citet_usename_uppercase:n {
  \prop_get:NnNTF \l__citet_names_prop {#1} \l_tmpa_tl {
    \text_titlecase_first:n \l_tmpa_tl
  } {
    \PackageWarning{citet}{Citation~key~#1~was~not~found.}
    \textbf{??}
  }
}

\NewDocumentCommand{\addcite}{mm}{
  \citet_addname:nn {#1} {#2}
}

\NewDocumentCommand{\citet}{m}{
  \citet_usename:n {#1} ~ \cite{#1}
}

\NewDocumentCommand{\Citet}{m}{
  \citet_usename_uppercase:n {#1} ~ \cite{#1}
}

\NewDocumentCommand{\citeauthor}{m}{
  \citet_usename:n {#1}
}

\NewDocumentCommand{\Citeauthor}{m}{
  \citet_usename_uppercase:n {#1}
}

\ExplSyntaxOff
\addcite{Bartram_1995}{Bartram et al.}
\addcite{Baudisch_2003}{Baudisch and Rosenholtz}
\addcite{Bederson_1996}{Bederson et al.}
\addcite{Bederson_1999}{Bederson and Boltman}
\addcite{Bell_2014}{Bell et al.}
\addcite{Besancon_2017a}{Besançon and Dragicevic}
\addcite{Besancon_2019}{Besançon and Dragicevic}
\addcite{Brodkorb_2016a}{Brodkorb et al.}
\addcite{Card_1999}{Card et al.}
\addcite{Carpendale_2001}{Carpendale and Montagnese}
\addcite{Cockburn_2009}{Cockburn et al.}
\addcite{Cockburn_2020}{Cockburn et al.}
\addcite{Coltekin_2017b}{Çöltekin et al.}
\addcite{Cox_1946}{Cox}
\addcite{Cumming_2013a}{Cumming}
\addcite{Cybulski_2021}{Cybulski}
\addcite{Dragicevic_2016}{Dragicevic}
\addcite{Drocourt_2011}{Drocourt et al.}
\addcite{Elmqvist_2008}{Elmqvist et al.}
\addcite{Franke_2020}{Franke et al.}
\addcite{Fujita_2010}{Fujita et al.}
\addcite{Furnas_1986}{Furnas}
\addcite{Furnas_1995}{Furnas and Bederson}
\addcite{Ghani_2011}{Ghani et al.}
\addcite{Gustafson_2008}{Gustafson et al.}
\addcite{Harrower_2007}{Harrower and Sheesley}
\addcite{Heer_2007}{Heer and Robertson}
\addcite{Jackle_2016}{Jäckle et al.}
\addcite{Javed_2012a}{Javed et al.}
\addcite{Javed_2012b}{Javed et al.}
\addcite{Jusufi_2011}{Jusufi et al.}
\addcite{Karnick_2010}{Karnick et al.}
\addcite{Krzywinski_2013}{Krzywinski and Altman}
\addcite{Lekschas_2020}{Lekschas et al.}
\addcite{Li_2023}{Li et al.}
\addcite{Purchase_2002}{Purchase}
\addcite{Reach_2017}{Reach and North}
\addcite{Reach_2018}{Reach and North}
\addcite{Reckziegel_2021}{Reckziegel et al.}
\addcite{Ruddle_2016}{Ruddle et al.}
\addcite{Scharr_2000}{Scharr}
\addcite{Shanmugasundaram_2008}{Shanmugasundaram and Irani}
\addcite{Skopeliti_2013}{Skopeliti and Tsoulos}
\addcite{Slocum_2014}{Slocum et al.}
\addcite{Snyder_1997}{Snyder}
\addcite{Tobler_1970}{Tobler}
\addcite{Tong_2018}{Tong et al.}
\addcite{Treves_2012}{Treves and Bailey}
\addcite{Treves_2017}{Treves and Skarlatidou}
\addcite{Tversky_2002}{Tversky et al.}
\addcite{Zhou_2018}{Zhou et al.}
\addcite{vanHam_2009}{{van Ham} and Perer}
\addcite{vanWijk_2003}{van Wijk and Nuij}
\addcite{vanWijk_2004}{van Wijk and Nuij}
\addcite{vonLandesberger_2016}{{von Landesberger} et al.}
\addcite{OSF-Transitions}{Hirsch et al.}
\addcite{OSF-Transitions}{Franke et al.}
\addcite{OSF-Transitions}{Franke}

\usepackage[nolist,nohyperlinks]{acronym}
\acrodef{tpeqd}[\textsc{tpeqd}]{two-point equidistant projection}
\acrodef{azeqd}[\textsc{azeqd}]{azimuthal equidistant projection}
\acrodef{CI}[CI]{confidence interval}
\acrodef{DH}[DH]{digital humanities}
\acrodef{DoI}[DoI]{degree-of-interest}

\onlineid{0}

\vgtccategory{Research}

\vgtcinsertpkg

\title{Two-point Equidistant Projection and Degree-of-interest Filtering for Smooth Exploration of Geo-referenced Networks}

\author{Max Franke\thanks{e-mail: \emph{\{first name\}.\{last name\}}@vis.uni-stuttgart.de}
\and Samuel Beck\footnotemark[1]
\and Steffen Koch\footnotemark[1]}
\affiliation{\scriptsize University of Stuttgart, Germany}

\teaser{
  \centering
  \includegraphics[height=6cm]{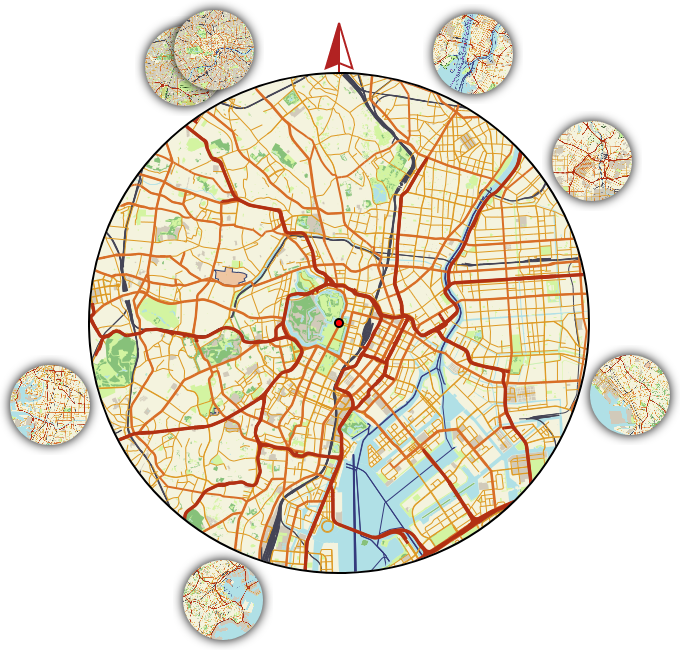}%
  \hfill%
  \includegraphics[height=6cm]{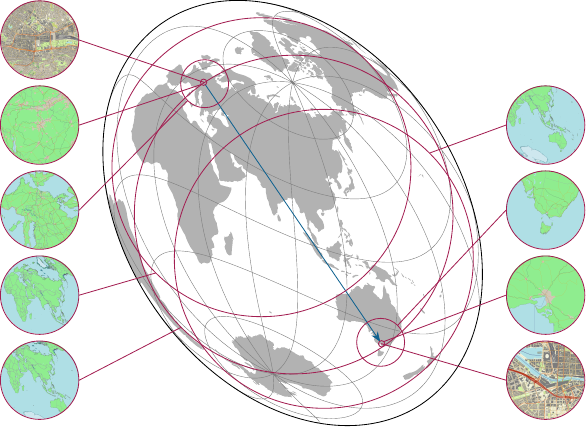}%
  \caption{
    Our prototype (left) shows an ego-perspective of a geo-referenced network via a large, central map of the focused node.
    Neighboring nodes are indicated by small proxy maps at the edge.
    Clicking a proxy starts an animated zoom-and-pan transition~\cite{vanWijk_2003} to navigate to the respective node.
    Stages along this transition, which is realized as a \acl{tpeqd}, are indicated on the right.
    The geodetic line (blue arrow) between start and end node is projected without distortion.
  }%
  \label{fig:teaser}%
}

\abstract{
The visualization and interactive exploration of geo-referenced networks poses challenges if the network's nodes are not evenly distributed.
Our approach proposes new ways of realizing animated transitions for exploring such networks from an ego-perspective.
We aim to reduce the required screen estate while maintaining the viewers' mental map of distances and directions.
A preliminary study provides first insights of the comprehensiveness of animated geographic transitions regarding directional relationships between start and end point in different projections.
Two use cases showcase how ego-perspective graph exploration can be supported using less screen space than previous approaches.
} % end of abstract


\hypersetup{
  colorlinks=false,  % do not color links
  breaklinks,        % break URLs (e.g., in references)
}
\PassOptionsToPackage{hyphens}{url}

\ifpdf%
  \hypersetup{
    pdfversion=1.4
  }
\fi

\begin{document}


\firstsection{Introduction}%
\iffalse\section{Introduction}\fi%  for VSCode structure
\label{sec:intro}

\maketitle

Geo-referenced networks, where vertices are associated with geographical locations, are relevant for many domain research questions.
Examples include transit networks, supply chains in global trade, or stops and related points of interest in the travel itinerary of a historical figure.
Map-based visualization approaches help to understand such geo-referenced graphs and often support their interactive exploration.
Since node positions are associated with geographic places, layout algorithms improving legibility and aesthetics~\cite{Purchase_2002} cannot be applied in a straightforward way.
As a consequence, the legibility of such graphs can quickly deteriorate when positions are distributed heterogeneously;
for example, with locally dense clusters that are globally connected.
A global overview of such graphs often displays large empty map regions to maintain the geographical distances.
Additionally, local details cannot be discerned easily without zooming in, which can lead to losing the global context.
Many approaches dealing with these problems pursue overview+\allowbreak{}detail, focus+\allowbreak{}context, or cue-based strategies~\cite{Cockburn_2009}.
Some of these works~\cite{Ghani_2011,Baudisch_2003,Gustafson_2008} consider an ego-perspective display of the graph, where a central map shows one vertex in larger detail, and connected vertices are indicated along the border.
These representations are more suitable for use cases where not the entire graph is of interest but rather the interactive exploration of a sub-graph and its immediate neighbors.
To support the interactive graph exploration while maintaining the understanding of scale and direction, animated transitions between vertices can be used~\cite{vanWijk_2003,Tversky_2002,Bederson_1999}.
Our approach falls into this category and in particular aims at improving the understanding of spatial properties of the graph during transitions.
Encoding the information about (large) distances and directions into animated transition opens up possibilities to show more space-efficient representations of geo-referenced sub-graphs.

\begin{enumerate*}[label=(\arabic*)]
  \item We combine different map projections with animated transitions to support exploring geo-referenced graphs and compare how well relative direction between locations can be understood during animation in a preliminary user study.
  \item We extend these animated transitions to consider additional points of interest including context clues, which support planning subsequent graph exploration steps.
    This functionality is showcased with two different prototypes using Mercator and \acl{tpeqd}.
\end{enumerate*}

\section{Related Work}%
\label{sec:rw}

Our work is related to previous publications that explore interactive or animated navigation through large spaces, where different subsets of interest are best viewed at vastly different scales.
\paragraph{Exploration of geo-referenced networks.}
\Citet{Carpendale_2001} suggest a united framework for presentation methods that unite an overview of the data space with smaller regions of higher detail.
Besides seamless distortions within the overview space, they suggest the term \emph{\enquote{inset}} for delimited regions of higher detail that occlude the space they magnify in the overview space.
For such regions that are spatially offset from the region they magnify and connected to it via lines or other visual indicators, they use the term \emph{\enquote{offset.}}
We use the terminology of \citeauthor{Carpendale_2001}, alongside the term \emph{\enquote{proxy}} for magnified regions that indicate regions outside the shown overview space.
Early work on exploration in large documents~\cite{Bederson_1996} and graph visualizations~\cite{Bartram_1995} utilizes insets or offsets with a higher level of detail to show regions of interest.
\Citet{Karnick_2010} use inset maps to show turns in higher detail on an overview map of a route.
\Citet{Lekschas_2020} use clustered offsets to highlight details in large data spaces such as maps and gigapixel images.
Many other works use offsets to magnify regions of interest in maps~\cite{Javed_2012a,Brodkorb_2016a}, or distort space to magnify such regions~\cite{Elmqvist_2008}.
%
% ego-perspective: proxies
The above works using insets or offsets are generally overview+\allowbreak{}detail approaches~\cite{Cockburn_2009}.
On the other hand, focus+\allowbreak{}context approaches show one region of interest in a larger view, and indicate off-screen regions of interest in proxies around it.
Early work, such as \emph{Halo}~\cite{Baudisch_2003} and \emph{Wedge}~\cite{Gustafson_2008}, show indicators without data content.
Later approaches~\cite{Jusufi_2011,Ghani_2011,Franke_2020} also show the contents of off-screen regions of interest.
The work of \citet{Ghani_2011} is closely related to our work regarding proxy maps and ego-perspective graph exploration.
However, their approach does not use zoom-and-pan transitions to give a spatial overview, and was designed with geographically localized, homogeneously distributed vertices in mind.
Our approach also extends earlier work by combining proxy maps with \ac{DoI} filtering~\cite{Furnas_1986,vanHam_2009} to indicate the most relevant off-screen regions of interest during transitions.

% zoom-and-pan animated transitions/navigation
\paragraph{Animated map transitions.}
Animated, continuous transitions between two visual states have been shown to facilitate understanding by letting users keep their mental model intact~\cite{Tversky_2002,Heer_2007,Bederson_1999,Shanmugasundaram_2008}.
\Citet{Treves_2017} suggest ideal paths of zoom and pan for animated map transitions where the spatial relationship of two or more locations is shown.
They recommend that the entire path including start and end point should be visible at once at some point during the transition.
\Citet{vanWijk_2003} give concrete recommendations for the zooming and panning functions over time for a smooth transition.
Their optimal path also has both the start and end point visible at the most zoomed-out position.
\Citet{Reach_2018} extend the work of \citeauthor{vanWijk_2003} with smoothing and easing recommendations, and interruptibility.
Other works explored tilting in zoom-and-pan transitions~\cite{Fujita_2010}, transitions following visual salience~\cite{Javed_2012b}, and the influence of three-dimensional maps~\cite{Harrower_2007,Li_2023} and abstract versus realistic map material~\cite{Cybulski_2021,Coltekin_2017b} on viewers' understanding of the relationships between shown places.

For animated transitions and scale-space navigation in geographical data, a suitable projection~\cite{Snyder_1997,Slocum_2014} from geographical space to two-dimensional Cartesian space is essential.
While some approaches utilize a three-dimensional representation of the world, information visualization approaches typically work better in combination with two-dimensional data representation spaces, and require less resources.
Mercator projection is often used, in particular in web-based geographical visualization approaches.
However, Mercator projection distorts areas towards the poles strongly, and different projections are required, for example, in the Arctic~\cite{Drocourt_2011,Skopeliti_2013}.
In addition, the direction towards distant locations differ between \enquote{how the crow flies} (geodetic line) and the constant-bearing, straight-line connection in Mercator projection (loxodrome).
Our approach utilizes \ac{tpeqd}~\cite{Cox_1946}, which projects Earth based on two arbitrary locations.
The surroundings of the geodetic line connecting these two locations are projected with few distortions.
We combine \ac{tpeqd} with animated zoom-and-pan transitions between the two locations for a more faithful representation of scale, direction, and distance.

\section{Approach}%
\label{sec:approach}

\begin{figure}[tb]
  \centering
  \includegraphics[width=\columnwidth]{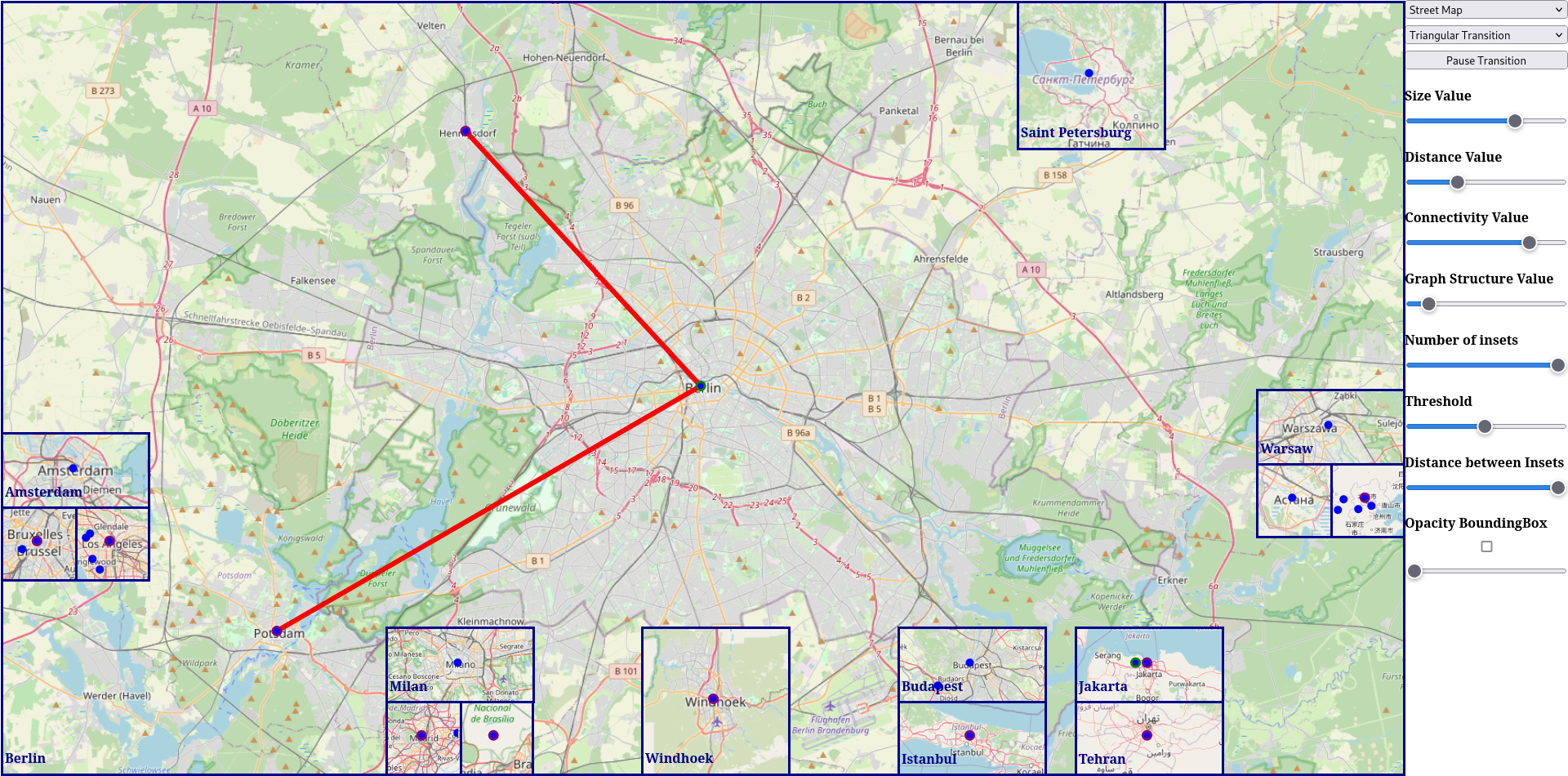}
  \caption{
    Our initial prototype showed maps in Mercator projection.
    On-screen vertices are connected explicitly.
    Proxies to off-screen vertices are shown at the edge, and aggregated as needed.
    Various \ac{DoI} functions can be combined with different weights for a total \ac{DoI} function that decides which vertices to show.
  }%
  \label{fig:prototype1}%
  \vspace{-4mm}
\end{figure}

Our approach (see \cref{fig:teaser,fig:prototype1}) can be seen as an extension of the concept presented by \citet{Ghani_2011}.
A large central map shows a geographical ego-perspective of the data.
In the static case, this is centered on a point of interest in the data (i.e., a vertex in the geo-referenced graph).
Related points of interest (e.g., neighboring vertices in the graph) are marked in the large map if they lie within its viewport.
Otherwise, they are indicated by small proxy maps at the edge in the direction of their off-screen position, as shown in \cref{fig:teaser}.
Users can navigate through the graph edge by edge through clicking on any of the offered neighboring vertices.
This initiates a smooth zoom-and-pan transition, as presented by \citet{vanWijk_2003}.
During the transition, the central map zooms out and pans towards the new vertex, and at one point, both vertices are visible at the same time (see \cref{fig:teaser}, right).
The other points of interest are updated in position, and can move between proxies and markers in the central map depending on the current viewport.
When the transition starts, vertices of interest to the destination vertex are shown as well, so that users can already plan their next hop.

\paragraph*{Map projection.}
The direction to off-screen points of interest and the zoom-and-pan transition happen in the projected, two-dimensional Cartesian space.
The choice of map projection influences how well spatial relationships, such as distance and direction, can be communicated in an animated transition.
Mercator projection has a clear advantage of efficiency in rendering:
Pre-rendered map tiles from tile servers can be used and only need to be scaled and translated.
This has led to the proliferation of this projection in interactive maps, especially on the web.
As a consequence, users are generally familiar with this projection, which is also an advantage.
However, Mercator projection has several drawbacks as well:
Areas towards the poles are represented disproportionately large.
In addition, geodetic lines are in general not straight lines when projected into Mercator.
Obviously, the choice of projection is also dependent on the application domain and use case~\cite{Snyder_1997,Slocum_2014}.
In our experiments, we found that animated transitions between two geographical locations should follow a straight line in the image space (i.e., the projected space).
Discussions with domain experts in various \ac{DH} collaborations (e.g.,~\cite{Franke_2020}), as well as later experiments, indicate that using Great Circle direction (i.e., along geodetic lines) can greatly facilitate the understanding of relative directions.
These observations indicate that using a map projection where the geodetic line between start and end position is also a straight line in the projected space can be advantageous for communicating relative direction and distance between two places.

Because of its popularity, and ease of implementation, our first prototype (see \cref{fig:prototype1}) uses Mercator projection.
To ensure a smooth zoom-and-pan transition, we precompute the transition path and fetch the required tiles~\cite{github/map-transition-helper} before the transition starts.
However, over longer distances, the straight-line path in the projection space deviates from the shortest path, and both direction and scale might not be communicated adequately.

We, therefore, created a second prototype (see \cref{fig:teaser}), in which \acf{tpeqd} is used.
\Ac{tpeqd} projects Earth based on two arbitrary locations, for which we choose the start and end point of the individual transition.
All distances from these two locations are represented without distortion.
As a consequence, the straight-line connection between them, which is also the geodetic line, is projected without distortion, and the area surrounding the points and the line is virtually undistorted.
This is extremely useful for a zoom-and-pan transition.
For far-apart projection nodes in \ac{tpeqd}, as well as locations far away from the projection node of \ac{azeqd}, shapes and areas get distorted.
We have found that familiar coastlines and country outlines are difficult to recognize further away than \SI{75}{\%} of the way from the projection point to its antipode, or about \SI{15000}{\kilo\meter}, in \ac{azeqd}.
The same holds for locations far from the central projection line in \ac{tpeqd} if the projection nodes are further apart than that.
A drawback of \ac{tpeqd} is that it needs to be recalculated for each pair of start and end points.
Our prototype does this in advance, using vector map data from NaturalEarth and OpenStreetMap to render the map data for each possible transition~\cite{github/tpeqd-rendered-transitions}.
Our own experiments, and discussions with domain experts in the \ac{DH}, revealed that (at least in the western hemisphere) users intuitively expect north to be up.
We rotate the projection space to ensure that north is up at the midpoint between start and end (see \cref{fig:teaser}, right).
For the static, zoomed-in view on single vertices, we use \ac{tpeqd} with both projection nodes in the position of the vertex (i.e., \ac{azeqd}).
Here, north is up as well.
At the start of the transition, we first transition the second projection node to the destination vertex, which leads to an initial rotation of the map centered on the start vertex.
Different azimuths at the projection nodes are unavoidable;
we draw an arrow pointing north to try to counteract the loss of orientation by viewers.
At the end of the transition, the first projection node is similarly transitioned to the destination vertex.

\paragraph*{Degree-of-interest filtering.}
To improve scalability, and to offer ways to include domain-specific requirements, we also introduced an adaptable \acf{DoI}~\cite{Furnas_1986} rating system.
Each vertex in the graph is assigned a \ac{DoI} score between zero and one.
Our prototype offers a set of \ac{DoI} functions that can be combined and weighted (see \cref{fig:prototype1}).
We offer separate measures based on the geographical distance, topological distance in the graph from the current vertex, vertex degree, as well as the area and the population of the cities (i.e., the vertices) as examples.
For an application in a specific domain, measures that are relevant to the concrete research questions at hand could easily be added here.
Assigning a \ac{DoI} score to every vertex has the advantage that a representative sample of proxy maps can be shown, instead of aggregating or displacing proxies;
for example, the $\le n$ proxy maps with a \ac{DoI} score above a configurable threshold can be shown, for an arbitrary $n$.

Our concept also extends the ego-perspective by adapting the set of edges in the visualized graph:
While the original edges are used for the graph-specific measures (topological distance and vertex degree), the shown proxy maps are determined by the \ac{DoI} score.
If a vertex' \ac{DoI} score is high, it might be shown even if it is not a direct neighbor of the currently focused vertex.
This concept becomes powerful when combined with domain-specific \ac{DoI} measures.
For instance, if the original graph represents the itinerary of a historical person's journey, that journey's path (i.e., the geo-referenced graph itself) is interesting to analyze.
It could also be interesting to see to which persons, in which cities, the historical person would send letters from each stop of the journey.
Or, which persons, from which cities, they met along the way.
We believe that combining \ac{DoI} functions, ego-perspective exploration of geo-referenced graphs, and domain-specific data is generalizable to many fields and problems.

\section{Evaluation}%
\label{sec:evaluation}

\begin{figure}[tb]
  \centering
  \begin{subfigure}{150pt}
    \includegraphics[height=95pt]{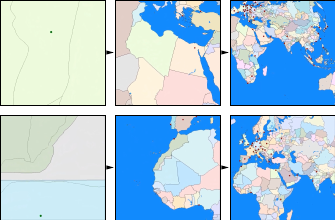}%
    \caption{Example stimulus frames}%
    \label{fig:stimuli:frames}
  \end{subfigure}
  \hfill%
  \begin{subfigure}{70pt}
    \includegraphics[width=70pt]{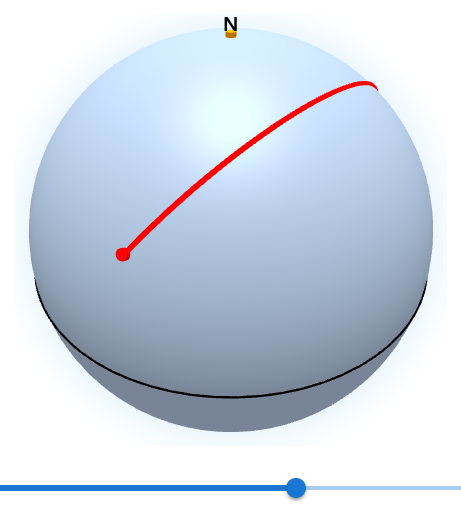}%
    \caption{Direction task}%
    \label{fig:stimuli:task}
  \end{subfigure}

  \caption{
    Example stimuli frames~(a) for the \ac{tpeqd} projection~(top) and Mercator projection~(bottom).
    The first half, until the point where both locations are visible, of two animated transitions are shown.
    The participants then had to specify the direction from the start point to the end point on the transition on an empty globe~(b).
  }%
  \label{fig:stimuli}%
  \vspace{-2mm}
\end{figure}

\begin{figure}[tb]
  \centering
  \includegraphics[width=\columnwidth]{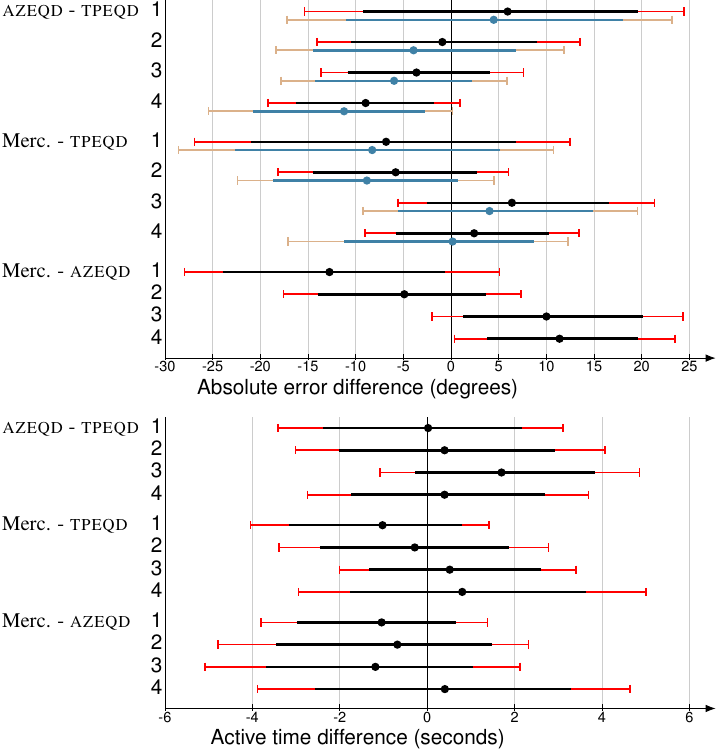}
  \caption{
    \Acp{CI} of the pairwise differences between the three projections, with Bonferroni correction (factor of 9) indicated by red whiskers.
    The \acp{CI} are calculated for the absolute error in angle of the direction task, and for the active time.
    For the error differences involving \ac{tpeqd}, the estimation of the corrected error when rotating the projection such that north is up for the start point (see discussion at the end of \cref{sec:evaluation}) is plotted in blue.
  }%
  \label{fig:pairwise-differences}%
  \vspace{-2mm}
\end{figure}

To explore what projection would be most suitable for animated transitions, we designed a quantitative user study, which we pre-registered~\cite{OSF-Transitions} on OSF.
In the study, we compared three map projections regarding their suitability to communicate relative direction:
\Acf{tpeqd}, \acf{azeqd}, and Mercator projection.
Despite its shortcomings, Mercator projection should still be considered for animated point-to-point transitions because of its familiarity and the efficient pre-generation and retrieval of map material.
Many other choices for geographical projections exist~\cite{Snyder_1997}, all with their individual advantages and drawbacks.
For our analysis, we considered two choices (\ac{tpeqd} and \ac{azeqd}) based on two requirements:
\begin{enumerate*}[label=(\arabic*)]
  \item Distances should be represented without distortion from the start point of the transition, and
  \item the straight-line connection between start and end point of the transition should be a geodetic line.
\end{enumerate*}
\Ac{tpeqd} and \ac{azeqd} seemed like good choices here, see \cref{sec:approach}.
In our study, we chose the start point of the transition as the projection node for \ac{azeqd}, but the end point could be chosen as well if appropriate for the concrete use case.

\paragraph*{Study design.}
We performed a quantitative study comparing the three projections (Mercator, \ac{tpeqd}, and \ac{azeqd}) online, using the Prolific platform.
Participants were not pre-screened, but we asked to participate only from a laptop or desktop computer, and added a scaling task to ensure all participants saw the stimuli in the same size.
The participants were chiefly male (45 of 72), and most were between 18 and 34 years old (59 of 72).
In the study, participants were presented with video stimuli (see \cref{fig:stimuli:frames}) showing an animated zoom-and-pan transition, using the technique of \citet{vanWijk_2003} with $\rho=1.4$, in one of the three projections.
The study design consisted of two independent variables: the projections, and the distance between the start and end point picked randomly from four difficulty intervals:
\begin{enumerate*}[label=(\arabic*)]
  \item 500--\SI{3000}{\kilo\meter},
  \item 3000--\SI{6000}{\kilo\meter},
  \item 6000--\SI{9000}{\kilo\meter}, and
  \item 9000--\SI{12000}{\kilo\meter}.
\end{enumerate*}

With a mixed design, the projection types were distributed \emph{between-subject,} while the distance was varied \emph{within-subject.}
Afterwards, participants were asked to specify the direction towards the end point on a three-dimensional globe (see \cref{fig:stimuli:task}).
After noticing that participants took shortcuts and memorized the geographical location of the end point in a preliminary study, we decided not to show any geographical features except for the start point, poles, and equator on the globe.
With \num{10} repetitions per condition, each participant was shown \num{40} stimuli during the main study.
We measured
the total time between the start of the task and the participant submitting their answer,
the \emph{active} time between the first movement of the mouse and the submission, and
the angular difference (error) between their answer and the correct azimuth.

\paragraph*{Evaluation results.}
In total, the data of \num{72} participants could be used for evaluation.
Of these, \num{21} were assigned to the Mercator projection conditions, \num{25} to the \ac{tpeqd} conditions, and \num{26} to the \ac{azeqd} conditions.
We carried out a time and error analysis with the sample mean per participant and condition, using interval estimation with \SI{95}{\%} \aclp{CI}, following the recommendations in recent work~\cite{Cumming_2013a,Besancon_2017a}.
The intervals are estimated using bias-corrected and accelerated ($BC_a$) bootstrapping with \num{10000} iterations.
We also calculate pairwise differences (see \cref{fig:pairwise-differences}) between the projections per difficulty, where we perform Bonferroni correction for multiple comparisons with a factor of \num{9}.

The pairwise differences show few conclusive results.
However, there is weak evidence for \ac{azeqd} having lower error rates at larger distances (level 4) than \ac{tpeqd}, and a trend towards the same conclusion for lower, increasing distance levels.
In addition, there is a trend for Mercator projection performing better regarding error rates for lower distances than both \ac{azeqd} and \ac{tpeqd}, but performing worse for distance levels 3 and 4.
For \ac{azeqd}, there is even weak evidence that it is better than Mercator projection for distance level 4.
Regarding time, there is no conclusive evidence, but some weak trends for Mercator projection being quicker for small distances, and \ac{azeqd} and \ac{tpeqd} being quicker for larger distances.

Participants rated the direction task as being quite hard in general, but especially so for \ac{azeqd}.
The study design exhibits some issues, without which the results might have been clearer.
For one, participants seemed to struggle with the unfamiliar distortions of \ac{azeqd} and \ac{tpeqd}.
However, Mercator projection, while being more familiar, also contains distortions many people are not aware of.
A qualitative think-aloud study analyzing participants' thought processes during geographical exploration in differently projected maps could be an interesting direction for future research.
Another issue is the direction task.
Here, participants were asked for the bearing from the start point, which might have skewed results slightly towards Mercator.
In Mercator projection, the transition followed the loxodrome (i.e., along a constant bearing), while for \ac{azeqd} and \ac{tpeqd}, it followed a Great Circle arc.
Hence, the bearing is not constant over the transition in those latter projections.
It is not clear how that affected the results, and it is challenging to validate how participants intuitively think about direction on a globe.
In terms of the direction, the rotation of the projections for \ac{azeqd} and \ac{tpeqd} might have impacted the difficulty of the task.
Here, the stimuli had different azimuths for the start and end point, which could affect how participants understood the directional relationship.
For \ac{tpeqd}, the question arises whether rotating the projection such that north is up at the start point would not be more intuitive, given the task.
We estimated that change by correcting the direction answers of that group by the difference between the azimuth at the start and the middle of the transition.
The performance for \ac{tpeqd} increased slightly both for correctness and for speed (see \cref{fig:pairwise-differences}, blue \acp{CI}).
Finally, the stimuli generation ensured a path with sufficiently many landmarks, which skewed the transitions towards land-covered areas closer to the equator.
This might have favored Mercator projection, which should perform worse in areas closer to the poles.
In total, the evaluation indicates that \ac{tpeqd} and \ac{azeqd} are promising for zoom-and-pan transitions over longer distances.

\section{Discussion and Outlook}%
\label{sec:discussion}

We consider our approach a promising improvement for the interactive exploration of geo-referenced networks.
Our initial prototypes use a pre-defined graph, but this could be extended to general geo-referenced data.
In that case, the \ac{DoI} functions would determine what other vertices to show depending on the current viewpoint and other parameters.
We already explored this with the first prototype (Mercator projection, see \cref{fig:prototype1}), with a weighted set of \ac{DoI} functions including geographical distance and city population.
For datasets from a concrete domain, with specific research questions, more individual \ac{DoI} functions could be formulated.
%The initial prototype shows that our approach works in Mercator projection as well.
Future work will compare the comprehensibility of the animated transitions, and also evaluate their interplay with the \ac{DoI} filtering options.

Our evaluation results of the animated transitions reveal that Mercator projection performs worse than \ac{tpeqd} and \ac{azeqd} for longer transitions.
Considering the technical advantages of precomputed WebMercator map tiles, a hybrid approach that uses Mercator projection for zoomed-in areas and \ac{tpeqd} for the zoomed-out parts of animated transitions could be an option.
Our experiments have shown that for the zoomed-in parts of the map around the undistorted geodetic line between the \ac{tpeqd} projection nodes, Mercator-projected material is a near pixel-perfect replacement.
Improved data structures and caching could potentially support ad-hoc generation of \ac{tpeqd} transitions instead of using precomputed solutions.

Our evaluation revealed interesting connections between prior geographical knowledge and sense of scale and distance.
Future studies could compare well-known with unknown map material, such as maps of unfamiliar local regions~\cite{Treves_2017} or other planets~\cite{vanWijk_2003}.
Another promising direction of future work is to augment the smooth transition path with points of interest along the way.

\acknowledgments{
  We thank Markus Stengel and Alexandra Hirsch for their contributions.
  This work has been partially funded by the EU Horizon 2020 project InTaVia (grant agreement no.\,101004825).
  The map data shown in \cref{fig:teaser,fig:prototype1,fig:stimuli} is in part \textcopyright\ OpenStreetMap contributors.
}

\bibliographystyle{abbrv-doi-narrow}

\bibliography{references}

\end{document}